# What we have (not)learned from the ultrarelativistic heavy ion collisions

Guy Paic

*Instituto de Ciencias Nucleares, Universidad Nacional Autónoma de México*

**Abstract.** The field of ultrarelativistic heavy ion collisions is today a flourishing activity both on the experimental and on the theoretical side. Although the theoretical justifications to study these collisions was given already more than three decades ago and the experimental studies have a history of more than 25 years we are still very much in the dark as to the details of the processes and of the characteristics of the matter created in collisions. Increasing the energy of collisions has brought new insights but has also resulted with new challenges. In the present paper I will try from a personal perspective to report on the answers we have collected and on the problems we are faced with. The account is partial, taking into account that it is impossible to render justice to every aspect of the field.

## INTRODUCTION

The study of ultrarelativistic heavy ion collisions has already a history of ~25 years with experiments spanning a large range of energies from the AGS accelerator at BNL (5 GeV/c), the CERN heavy ion beams at the SPS accelerator (17 GeV/c) and finally the Relativistic Heavy ion Collider (RHIC) at BNL where the energies reach 200 GeV in the nucleon - nucleon center of mass.
The initial impetus for this study was to observe in the laboratory a phase transition between the ordinary hadronic matter to a deconfined plasma of quarks and gluons named the quark gluon plasma (QGP) by Shuryak [1]

From thermodynamical considerations, and from models based on the fundamental theory for the strong interaction, Quantum Chromo Dynamics (e.g. lattice QCD calculations), estimates for the critical temperature and the order of the transition can be made. Calculations indicate that the critical temperature should be $T_c = 192(7)$ (4) MeV, where the first number in parentheses refers to the statistical and the second to the estimated systematic error [2]. The order of the transition at various values of the chemical potential is not known. In general, a decreasing critical temperature with increasing chemical potential is expected. Likewise, at non-zero chemical potential a mixed phase of coexisting hadron gas, and QGP is predicted to exist in a certain temperature interval around the critical temperature. Recently, calculation techniques have progressed to the point of allowing an extension of the lattice methods also to finite chemical potential. Such calculations also suggest the existence of a critical point at larger chemical potential above which, the transition may be of first order.

The highest energy nucleus-nucleus collision has been so far achieved at RHIC: 100 GeV per nucleon. This means that each incoming nucleus is contracted by a Lorentz factor γ ≈ 100: nuclei are thin pancakes colliding. The collision creates thousands of particles in a small volume. These particles interact. If these interactions are strong enough, the system may reach a state of local thermodynamic equilibrium. Equilibrium is at best local, certainly not global: a global equilibrium applies to a gas in a closed box, which stays there for a long time and becomes homogeneous.

The fundamental problem of studies in heavy-ion physics is to determine precisely whether the system created in the collision reaches a state of collectivity in which a local thermodynamic equilibrium is reached.

Assuming that an equilibrated system is indeed created one then tries to understand the footprints left by the hadronizing initial system in the final hadronic state. We will try to review in a critical manner the present state of the field pointing more to the disagreement or incongruities between theory and experimental results than to the "present" majority view. Here it is important to underline that in the history we were confronted with many changes of attitudes or consensus, because, unfortunately we are often carried not by irrefutable arguments but more by the necessity to explain a given body of data. In that vein probably the most significant example is the evolution from the idea of an ideal gas of partons based on the asymptotic freedom assumption to the revision of our beliefs to the QGP essentially being, not an ideal gas but rather a perfect fluid also called a strongly coupled quark gluon plasma. This revision was the fruit of the results obtained at RHIC (the collective behavior, the ideal hydro flow with low viscosity the suppression of jets etc).

## ENERGY DENSITY

The first objective of the search for the quark gluon plasma should be to establish whether the energy density created in the collisions surpass the theoretically predicted value which is ~1 GeV/fm$^3$. The value of the energy density is usually done following the Bjorken recipe at mid rapidity. The central rapidity region is approximately boost invariant as required by the Bjorken estimate. Under boost invariance assumption, the energy density of the central rapidity region in the collision zone at formation time τ can be estimated by the Bjorken energy density [3].

$$\varepsilon = \frac{dE_t}{dy}\frac{1}{A\tau},$$

Where $E_t$ is the total transverse energy, $A$ is the area of the overlap of the two colliding nuclei and τ is the thermalization time. Since we have access only to charged particles in general the above formula is amended as follows:

$$\frac{d<E_t>}{dy} \approx \frac{3}{2}(<m_t>\frac{dN}{dy})_{charged pions} + 2(<m_t>\frac{dN}{dy})_{charged kaons and protons, antiprotons}$$

For this exercise STAR [4] uses the average measured transverse momenta. The factors 3/2 and 2 compensate for the neutral particles.

The densities obtained this way are of $[5.2 \pm 0.4] \times \frac{1}{\tau}$ GeV/fm$^2$ at 200 GeV for the most central collisions.

This estimate is unfortunately very much constructed of unknowns. How to take the area into account? For instance take the rms radius or the Saxon-Wood that is $\sqrt{2}$ times smaller? The other unknown is the thermalization time. The estimates vary greatly from 0.6 fm/c to 0.2 fm/c. It is safe to assume that the calculated values of energy density represent an estimate which can vary easily by up to a factor 2. Nevertheless it seems that the value of for the most central collisions lies well above the values predicted by the lattice QCD results which are giving about 1 GeV/fm$^2$ for the energy density at the transition to QGP. However here there are two comments to be made. One is that for peripheral collisions with ~ 90 participants the measured density drops to about half of the central value. This may mean that in very peripheral collisions we may observe collisions where the energy density may be very close to the critical energy density. The second comment is that from the top SPS energy to ~ 20 times larger energies one observes a rise of only about 40%, just about equal to the value extracted by STAR at 62.4 GeV! This lets the argument about a strongly increased density rely mostly on the considerations of the thermalization time.

On the other hand although the nuclei are contracted to very small values due to the value of the Lorentz factor γ~200 one should not forget that the low momentum gluons will form a cloud around the squashed nuclei so that the time of interpenetration may be higher than the one now considered.

Finally, one should bear in mind that the extracted values represent only an average over the area studied. Since for instance in central collisions we collide two squashed spheres, the values at the center will be much higher than at the outer radius. The other problem is to understand whether we can tell from experimental data whether the system has reached local equilibrium? One should keep in mind that equilibrium is, at best, an approximation. Even if it turns out to give reasonable results, it is not the end of the story [5].

## PARTICLE PRODUCTION

The particle abundances have been reproduced with a large success in a wide variety of energies and particle species with a remarkable success using a simple chemical equilibrium model [6].

The model makes use of baryon and strangeness numbers. The free parameters of the model are the freeze out temperature and the strangeness and baryonic chemical potentials. Using for example the ratios of charged kaons, and protons and antiprotons simultaneously one gets

$K^-/K^+ = \exp[(-2\mu_B/3+2\mu_S)/T_{chem}]$ and $\bar{p}/p = \exp(-2\mu_B/T_{chem})$,

Where $\mu_B$ and $\mu_S$ are the baryonic and strangeness chemical potentials, respectively and $T_{chem}$ is the chemical freezout temperature.

The results of a detailed analysis of the production of particles over a large range of energy (62.4 – 200 GeV) in Au-Au collisions results in a puzzling conclusion; the extracted $T_{chem}$ stays within the error bars equal in magnitude and equal to the pp value i.e $T_{chem} \approx 156$ MeV. A more recent analysis of the available data yields a somewhat higher chemical freeze-out temperature of 177MeV [7].

Apparently nothing distinguishes in the thermal model the production of hadrons in pp and in heavier system. The latest increase in the chemical freeze out temperature goes in the right direction closer to the calculated critical deconfinement temperature of 192 MeV [2], the existing difference may perhaps point to a deeper reason. One should remember that the extracted temperature lies well under the latest estimates for the transition temperature in QCD which is of 192 MeV. Perhaps there is an intermediate regime between the QCD transition and freeze-out during which the system created in a heavy ion collision persists in a dense hadronic phase.

While successful to give a unified view of the data one should not forget that the hadronization most probably does not happen at once especially for different species and that the success can be a coincidence of many causes. Recently, a calculation of the probability to form colorless clusters was done starting with a set of free quarks. The results showed that the formation of colorless clusters of three quarks occurs with a sharp jump at a critical energy density while the production of colorless clusters of quark-antiquark occurs smoothly from the low to the high energy density domains. The authors interpret this as a quantitative difference in the production of baryons and mesons in function of energy density, indicating a fundamental difference in the hadronization of the plasma [8].

## THE RADIAL FLOW

One of the achievement of the experiments with heavy ions at the SPS has been the establishment of the "radial flow". The transverse mass $m_T^2 = (m^2 + p_T^2)$ spectra of produced particles are sensitive to the collision dynamics. The shape of the distributions is approximately exponential [$\propto \exp(m_T/T)$]. The presence of strong radial flow in Pb–Pb collisions at the top SPS energy was deduced from the systematics of experimental data suggesting approximately a linear increase of the inverse slope with the particle mass. This was observed for the first time by the NA44 experiment [9]. Such a behavior can be best understood as a collective expansion of an initially dense system, be it partonic or hadronic, or both. Hydrodynamical models (which assume an isentropic expansion by definition) describe this effect by a radial collective velocity field that grows towards the surface, with the surface (and average) velocity increasing over the entire course of the expansion. Thus the flow fraction of the average kinetic energy increases while the temperature falls steeply, both observables reaching a certain characteristic value that characterizes the stage where emission products decouple from rescattering. Within a given centrality bin the particle spectra are fitted simultaneously by the blast-wave model [10], which assumes a radially expanding thermal source. The fits [11] provide simultaneous information about the radial flow velocity ($\beta$) and the kinetic freeze-out temperature ($T_{kin}$) at final freeze-out. $T_{kin}$ and $\beta$ show very similar dependences as a function of the pseudorapidity d$N_{ch}$/d$\eta$ in both Cu+Cu and Au+Au collisions, evolving smoothly from the lowest (p+p) to the highest (central Au+Au) available multiplicities. $T_{kin}$ decreases with centrality and thus implies that the freeze-out occurs at a lower temperature in more central collisions, although, let us remind that, the chemical freezeout temperature as commented above stays remarkably stable. Essentially the kinetic freezout occurs at ~140 MeV at low multiplicities and falls below 100 MeV for the most central collisions while at the same time the collective radial flow increases from ~0.24 [12] for proton-proton collisions (!) to ~0.6c for the most central collisions.

The RHIC results are interesting because of the fact that even for pp one finds a collective flow. This is a new feature because Shuryak and Zhirov [13] did report "no flow" in ISR data. In the community the result for pp is sometimes discounted on effects of jets. But then it is not clear where does the radial flow start and where does the effects of jets end? It points out to the fact that we have to be very careful with the interpretation of the data. The variation of the chemical and kinetic freezeout over the range of energies from 1 GeV to 200 GeV shows that up to center of mass of ~10 GeV there is essentially no difference between the two temperatures as one would expect. This suggests that kinetic freeze-out happens relatively quickly after or concurrently with chemical freeze-out. On the other hand the chemical freezout temperature saturates around 150 MeV at the same center of mass energy of 10 GeV. However the final kinetic freeze out temperature gets ever lower as the energy of the collision raises and the multiplicity gets larger. It is important to point out that the latest STAR data show that there is virtually no change of the $T_{kin}$ for the most central collisions in Au+Au from 63 till 200 GeV. This result is somewhat puzzling because naively one would expect that to a higher multiplicity corresponds a larger time of kinetic decoupling hence lower temperatures. This finding is in line with results of momentum correlations that also do not exhibit an increase in transverse radius [14]

## THE BARYON PRODUCTION MYSTERY

The studies of particle production as a function of $p_t$, in the momentum regions where identification is possible at RHIC, exhibit the following behavior: the p/π+ and pbar/π− ratios increase with pt up to ~2 GeV/c and then start to decrease for higher $p_t$ in both pp and Au + Au [15] collisions, reaching a value which corresponds to the fragmentation value observed in e+e− collisions for quarks and gluons[16]. The spectra at $p_t < 2$ GeV/c have been observed to follow a $m_t$ and $x_T$ scaling, consistent with a transition between soft and hard processes at around $p_t \sim 2$ GeV/c. The surprise lies in the fact that one would expect a ratio that does not exceed the fragmentation value i.e. ~0.2 as observed in e+e− collisions, while in the experiment the ratio rises up to more than one!

In the literature two possible explanation are prominently put forward:
- the hydrodynamical approach [17] where one assumes a local thermal equilibrium of partonic/hadronic matter at an initial time, describing the space-time evolution of thermalized matter by solving the equations for energy-momentum conservation in the hydro picture. Another model, where the radial flow and the size of the system of emitting particles are taken into account was proposed in [18] and can describe the proton to pion ratio for different centralities.
- A large class of models called generically "coalescence" where the particle species ratios observed in the intermediate $p_T$ regime (2-6 GeV/c) of heavy ion collisions are explained by a collective production mechanism, namely recombination or coalescence [19]. In most coalescence models hadrons are assumed to form from essentially collinear partons. The parton overlap function is sometimes simply assumed to be a delta function, or at best in some cases small finite transverse widths have been used, assuming an $x_T$ distribution like one expects to see in the final state hadron, such that the partons do not have to undergo a change in momentum when forming a hadron. Although the coalescence models have been accepted, they do not provide a satisfactory response to many questions [20].

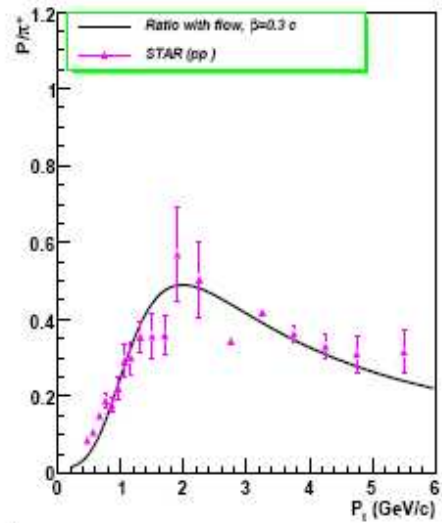

**FIGURE 1** Proton to pion ratio from pp collisions using HIJING, compared to pp data from STAR experiment. The solid line was obtained as the ratio of the fits to the simulated proton and pion spectra.

Recently [21], an attempt to fit all the existing data including into the event generator the radial flow component has successfully reproduced the features of the baryon/meson ratios from pp collisions to Au-Au ones, at different energies and centralities. In Fig.1 we show the quality of the fit obtained for pp collisions using the flow afterburner. Again the flow for pp collisions should be used in a cautious way. Most probably the minijets do have a contribution to the observed ratios. In any case I believe that the jury is still out on the various models on the market to explain the baryon meson puzzle until a thorough comparison is achieved for a large variety of data

## AZYMUTHAL FLOW

The azymuthal flow measures the asymmetry of particle density in momentum space relative to the reaction plane. The geometry is initially asymmetric in non-central A+A collisions. The transformation from geometric asymmetry to momentum-space asymmetry requires strong interactions at an early stage where the geometry has not been blurred by the expansion. Probably the most interesting finding is that the measured $v_2$ values as a function of $p_T$ scale with the number of constituent quarks in the measured hadron. This indicates that the number of the constituent quarks in a hadron is a relevant degree of freedom.

In addition, hadrons with strange quarks behave the same as other particles. The other important observation is that for the first time since the elliptic flow or azymuthal anisotropy has been measured starting at very low energies, the flow has reached values in agreement with what one would expect from hydrodynamic considerations [22]. There are however some aspects that merit mentioning. Nuclei colliding at ultrarelativistic energies have a large initial orbital angular momentum $L_0$ if their impact parameter is of order of some fm; in fact, for symmetric nuclei, $L_0 \sim A\sqrt{s_{NN}}b/2$ where b is the impact parameter [23]. At LHC the angular momentum will be almost two order of magnitude larger than at RHIC and will reach values of $L_0 \sim 1.4 \times 10^7$! Due to the inhomogeneity of the colliding nuclei in the transverse plane, a significant fraction of $L_0$ must be deposited in the interaction region, Large values of the initial angular momentum may enhance the elliptic flow and may lead to the polarization of the emitted particles. It is not clear to what extent these effects may be observed but including all the effects in a detailed manner seems important especially taking into account that lately calculations are extracting with viscous hydrodynamics the shear viscosity values of the formed matter [24] transforming thus the elliptic flow into a precision tool. The $v_2$ value that characterizes the elliptic flow is actually connected with the shear viscosity entropy ratio and this ratio may be represented by $\eta/s \approx T\lambda_f c_s$, where T is the temperature, $\lambda_f$ is the mean free path and $c_s$ is the speed of sound in the matter. The temperature T = 165 ± 3 MeV is constrained via a fit to the elliptic flow data $c_s$ = 0.35 ± 0.05, and $\lambda_f$= 0.3±0.03fm. This gives a value $\eta/s$= 0.09 ± 0.015 [25]. One is always somewhat surprised to see pure hydrodynamics being accepted as a proof of early thermalisation not considering the other effects that could contribute like the magnetic field effects.

# PARTONIC ENERGY LOSS

The hard jets created in hard parton scattering are produced swiftly after the collision and do not participate in the thermalization process that was considered in the parameters treated so far. The idea of Bjorken [J.D. Bjorkem Fermilab-pub 82/59-THY (1982)] was retaken with refined approaches in the 1990's [36]. The idea is that "implanting" a fast parton which traverses the dense matter and measuring the properties of the hadronic jet resulting from the hadronization one could gather information about the characteristics of the dense phase. Very early, the RHIC experiments reported a drastic reduction of the yield of hadrons [28]. The recorded yields above pt~ 4 GeV were by a factor of 5 smaller than the appropriately scaled proton proton values in central collisions [29]. It was farther verified that at more peripheral collisions the suppression was less as shown in figure 2. Later the d-Au collisions demonstrated that the effect was indeed due to final state interactions.

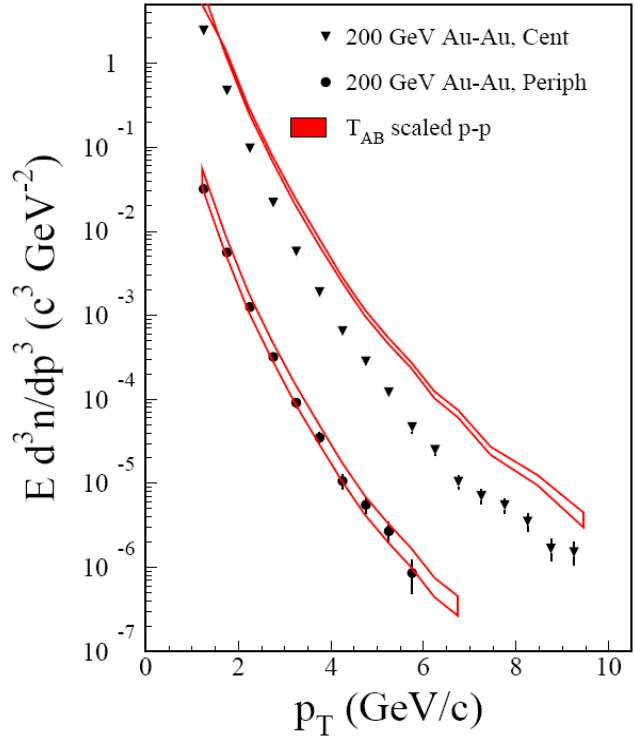

**FIGURE 2** $\pi^0$ $p_T$ spectra in 200 GeV Au+Au collisions from the PHENIX collaboration compared to a scaling of the 200 GeV p + p $\pi^0$ differential cross section . The central data were obtained with a 0–10% centrality cut while the peripheral data were obtained with an 80–92% cut. The data are from Phenix [29].

Another very important piece of experimental evidence for energy losses have been the dihadron azymuthal correlations. Although it is very difficult to detect jets at RHIC because of the low energ both STAR and PHENIX [30] have directly observed the presence of jets by studying two-hadron azimuthal-angle correlations. The pairs of particles are chosen such that one particle lies within a "trigger" $p_T$ range while the other "associated" particle falls within a lower $p_T$. In Fig.3 we show the first evidence was given by STAR.

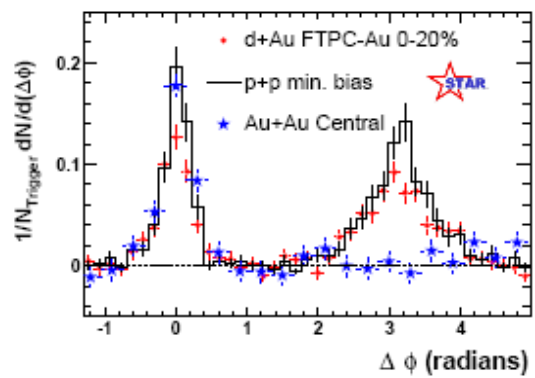

**FIGURE 3** Dihadron azimuthal correlations at high $p_T$ for p+p, central d+Au and central Au+Au collisions (background subtracted) from STAR.[31]

The suppression of the high $p_t$ spectra and the dihadron correlations have started a complete industry, both on the side of the experiments that have discovered new features like the "conical" emission or the ridge effects, into which we will not enter. Although the concept of energy loss is well known the implementation varies and we do not have yet a detailed view of the emission of partons out of the dense system. Additionally, it is not possible to draw a conclusion on the kind of matter the partons went through. Probably, what we have learned from the results so far is that the energy loss is larger than expected resulting that the surviving partons are emitted mostly from the surface of the system [32].

In spite of a huge experimental effort we are still not very advanced in the understanding of the parton energy loss. Two pillars of the community were recently put under scrutiny. One refers to the free transport of photons through the dense matter, hence a non suppression of the photon spectra in heavy ion collisions compared with the proto-proton case. At Quark matter 2008 [33] [34] the Phenix and tha STAR collaboration have reported on the apparent suppression of high momentum photons at momenta of ~6-7 GeV/c and reaching the same suppression as neutral pions at momenta of ~ 14-15 GeV/c. Of course the low statistics only wets our appetite for the time being but if it reveals true with higher statistics it would have a large impact on our vision of the parton energy loss as we understand it today.

The second pillar is the "dead cone effect" whereby the radiation of gluons from a massive parton is suppressed at angles $\theta < M_q/E_q$. Also the slower moving quark would meet on its way a more dilute medium (due to the expansion). Both effects should result in a reduced energy loss for heavy quarks [35]. However the RHIC experimental results do not support the picture! The semi-leptonic decays of mesons containing c and b quarks, in Au+Au collisions show a suppression very much alike the $\pi_0$'s above $p_t$ of 5 GeV/c. [36]

# THE HEAVY ION PHYSICS AT LHC

It is of course very tempting to make previsions for the LHC. Lately, a complete set of such predictions have been published [38]. The predictions do reflect the basic complexity of the field. Essentially the predictions are very different and only the experiments will be able to disentangle among them. One thing is certain: the hard processes will dominate at the LHC energies. The corollary of that is that the soft part of the spectrum will have an important contribution from the fragmentation. In the limit in which the fragmentation into different hadronic species differs from the results of the hadronization in the chemical equilibrium model, discussed above, one should take great care to subtract the fragmentation part before analyzing the soft part of the spectrum. The present "simple" taxonomy of the spectra into soft intermediate and high pt more or less well defined regions will no doubt suffer reappraisal.

The effect of the minijets and jets will seriously influence the measurements of flow. No doubt the physics of heavy ion collisions at the LHC will be exciting, but I believe that the path to the truth will be every inch as arduous as at RHIC energies.

# CONCLUSION

In this brief overview it was not possible to illustrate all the challenges of this very rich field. The ongoing experiments and do question the accepted truths daily and innovative approaches are constantly being tried. In this verve I would mention the recent attempts to study the effect of magnetic fields created in the collisions [39] and the possibility that some of the apparent difficulties pointed out in this paper can be explained taking into account that the multiplicity evolution of the soft portion of the $p_t$ spectra in collisions at RHIC could be dominated by phase-space restrictions more than fundamental physics [40].

# ACKNOWLEDGMENTS

The discussions with E. Cuautle and A. Ayala are gratefully acknowledged. The present work was supported by the contract DGAPA IN115808